\documentclass[twocolumn,tighten]{aastex62}
\usepackage{amsmath}
\usepackage{xspace}
\usepackage{datetime2}
\usepackage[utf8]{inputenc}

\newcommand{\keyw}[1]{\textcolor{gray}{#1}}

\defcitealias{Mathis:1977hp}{MRN}
\defcitealias{Pollack:1994fn}{P94}
\defcitealias{Birnstiel:2011ks}{B11}
\defcitealias{Birnstiel:2015cl}{B15}

\usepackage{savesym}
\savesymbol{tablenum}
\usepackage{siunitx}
\restoresymbol{SIX}{tablenum}

\graphicspath{{.}{figures/}}

\newcommand{\kabs}{\ensuremath{\kappa_\nu^{\mathrm{abs}}}\xspace}
\newcommand{\ksca}{\ensuremath{\kappa_\nu^{\mathrm{sca}}}\xspace}
\newcommand{\kscaeff}{\ensuremath{\kappa_\nu^{\mathrm{sca,eff}}}\xspace}
\newcommand{\kext}{\ensuremath{\kappa_\nu^{\mathrm{ext}}}\xspace}
\newcommand{\ktot}{\ensuremath{\kappa_\nu^{\mathrm{tot}}}\xspace}
\newcommand{\kabst}{\ensuremath{\kappa_\nu^{\mathrm{abs,tot}}}\xspace}
\newcommand{\kscat}{\ensuremath{\kappa_\nu^{\mathrm{sca,tot}}}\xspace}
\newcommand{\kextt}{\ensuremath{\kappa_\nu^{\mathrm{ext,tot}}}\xspace}
\newcommand{\amin}{\ensuremath{a_\mathrm{min}}\xspace}
\newcommand{\amax}{\ensuremath{a_\mathrm{max}}\xspace}
\newcommand{\rhos}{\ensuremath{\rho_\mathrm{s}}\xspace}

\definecolor{intrevcolor}{rgb}{0.0, 0.0, 1}

\newcommand{\paperdsharpisella}{Isella et al.\ ({2018})\xspace}

\newcommand{\paperdsharpandrewst}{Andrews et al.\ {2018}\xspace}
\newcommand{\paperdsharphuangringst}{Huang et al.\ {2018a}\xspace}
\newcommand{\paperdsharphuangspiralst}{Huang et al.\ {2018b}\xspace}
\newcommand{\paperdsharpisellat}{Isella et al.\ {2018}\xspace}
\newcommand{\paperdsharpguzmant}{Guzman et al.\ {2018}\xspace}

\newcommand{\paperdsharpdullemondt}{Dullemond et al.\ {2018}\xspace}

\submitjournal{ApJL}
\usepackage[capitalise,nameinlink]{cleveref}
\begin{document}

\title{The Disk Substructures at High Angular Resolution Project (DSHARP):\\V. Interpreting ALMA maps of protoplanetary disks in terms of a dust model}
\shorttitle{DSHARP V: Dust Model}
\shortauthors{Birnstiel et al.}

\correspondingauthor{Tilman Birnstiel}
\email{til.birnstiel@lmu.de}
\author[0000-0002-1899-8783]{Tilman Birnstiel}
\affiliation{University Observatory, Faculty of Physics, Ludwig-Maximilians-Universit\"at M\"unchen, Scheinerstr.~1, 81679 Munich, Germany}

\author[0000-0002-7078-5910]{Cornelis P.~Dullemond}
\affiliation{Zentrum f{\"u}r Astronomie, Heidelberg University, Albert Ueberle Str.~2, 69120 Heidelberg, Germany}

\author[0000-0003-3616-6822]{Zhaohuan Zhu}
\affiliation{Department of Physics and Astronomy, University of Nevada, Las Vegas, 4505 S.~Maryland Pkwy, Las Vegas, NV 89154, USA}

\author[0000-0003-2253-2270]{Sean~M.~Andrews}
\affiliation{Harvard-Smithsonian Center for Astrophysics, 60 Garden Street, Cambridge, MA 02138, USA}

\author[0000-0001-6906-9549]{Xue-Ning Bai}
\affiliation{Institute for Advanced Study and Tsinghua Center for Astrophysics, Tsinghua University, Beijing 100084, China}

\author[0000-0003-1526-7587]{David J.~Wilner}
\affiliation{Harvard-Smithsonian Center for Astrophysics, 60 Garden Street, Cambridge, MA 02138, USA}

\author[0000-0003-2251-0602]{John M.~Carpenter}
\affiliation{Joint ALMA Observatory, Avenida Alonso de C{\'o}rdova 3107, Vitacura, Santiago, Chile}

\author[0000-0001-6947-6072]{Jane Huang}
\affiliation{Harvard-Smithsonian Center for Astrophysics, 60 Garden Street, Cambridge, MA 02138, USA}

\author[0000-0001-8061-2207]{Andrea Isella}
\affiliation{Department of Physics and Astronomy, Rice University, 6100 Main Street, Houston, TX 77005, USA}

\author[0000-0002-7695-7605]{Myriam Benisty}
\affiliation{Unidad Mixta Internacional Franco-Chilena de Astronom\'{i}a, CNRS/INSU UMI 3386, Departamento de Astronom{\'i}a, Universidad de Chile, Camino El Observatorio 1515, Las Condes, Santiago, Chile}
\affiliation{Univ.~Grenoble Alpes, CNRS, IPAG, 38000 Grenoble, France}

\author[0000-0002-1199-9564]{Laura M.~P{\'e}rez}
\affiliation{Departamento de Astronom{\'i}a, Universidad de Chile, Camino El Observatorio 1515, Las Condes, Santiago, Chile}

\author[0000-0002-8537-9114]{Shangjia Zhang}
\affiliation{Department of Physics and Astronomy, University of Nevada, Las Vegas, 4505 S.~Maryland Pkwy, Las Vegas, NV 89154, USA}


\begin{abstract}
 The Disk Substructures at High Angular Resolution Project (DSHARP) is the
 largest homogeneous high-resolution ($\sim$0\farcs035, or $\sim \SI{5}{au}$)
 disk continuum imaging survey with ALMA so far. In the coming years, many more
 disks will be mapped with ALMA at similar resolution. Interpreting the results
 in terms of the properties and quantities of the emitting dusty material is,
 however, a very non-trivial task. This is in part due to the uncertainty in the
 dust opacities, an uncertainty which is not likely to be resolved any time soon.
 It is also partly due to the fact that, as the DSHARP survey has shown, these
 disk often contain regions of intermediate to high optical depth, even at
 millimeter wavelengths and at relatively large radius in the disk. This makes
 the interpretation challenging, in particular if the grains are large and have a
 large albedo. On the other hand, the highly structured features seen in the
 DSHARP survey, of which strong indications were already seen in earlier
 observations, provide a unique opportunity to study the dust growth and
 dynamics. To provide continuity within the DSHARP project, its follow-up
 projects, and projects by other teams interested in these data, we present here
 the methods and opacity choices used within the DSHARP collaboration to link the
 measured intensity $I_\nu$ to dust surface density $\Sigma_d$.
\end{abstract}

\section{Introduction}

Dust thermal (sub-)millimeter emission from the outer regions ($r\gtrsim
10\,\mathrm{au}$) of protoplanetary disks has traditionally been considered to
be optically thin, because it would require implausible amounts of dust mass to
make the disk optically thick at these wavelengths out to many tens of au. Even
if it were optically thick, it would produce much higher disk-integrated flux
values than are observed \citep{Ricci:2012ji}. If the assumption of low optical
depth were true, it would aid the interpretation of sub-millimeter continuum
maps in terms of the properties and dynamics of the dust grains, because the
observed intensity $I_\nu$ would be directly proportional to the underlying dust
surface density $\Sigma_d$.

Observational results from the past decade have shown that protoplanetary disks
do not have simple monotonically decreasing surface density profiles, but
consist of multiple narrow rings
\citep[e.g.,][]{ALMAPartnership:2015cg,Andrews:2016fq,Isella:2016hn,2017A&A...600A..72F,2018A&A...610A..24F,Huang:2018fl},
apparently single massive rings \citep[e.g.][]{Brown:2009hr,Casassus:2013ky} but
also often non-axisymmetric features such as lopsided rings
\citep[e.g.][]{vanderMarel:2013ky}, and spirals
\citep[e.g.][]{Perez:2016hu}. These narrow or compact features present
concentrations that may be optically thick or moderately optically thick. In
between these features, the material is often optically thin.

This is both a curse and a blessing. Such optical depth effects make the
interpretation of the data more difficult. In particular the spectral slope
variations at (sub-)millimeter wavelengths ($I_\nu \propto
\nu^{\alpha_\mathrm{mm}}$) across rings and gaps are strongly affected, perhaps
even dominated, by these effects. But optical depth effects also provide new
opportunities to measure the properties of the dust. An example of this is the
scattering of its own thermal emission, and the induced polarized millimeter
emission \citep{Kataoka:2015hl}. Another example is when dust rings extinct part
of the CO line emission from the back side of a disk (\paperdsharpisellat). 

But there is, of course, the major uncertainty in the dust opacity law. This is
a long-standing problem \citep{Beckwith:1991ce} that still has not been
resolved. This is in part because dust in protoplanetary disks, in particular
the large grains seen as settled grains in a thin mid-plane layer, is very
different from the dust in the interstellar medium. In part it is, however, also
due to uncertainties in the numerical and conceptual challenges in computing
opacities. When comparing computed opacities with laboratory measured
opacities in the millimeter range, one often sees discrepancies of up to a
factor of 10 or more \citep[e.g.][]{2017A&A...600A.123D}.

By inspecting the (sub-)millimeter spectral slope $\alpha_\mathrm{mm}$ we can
learn something about the grain size distribution and the opacity law
\citep[e.g.][]{Beckwith:1991ce,2003A&A...403..323T,Wilner:2005bu}. With high
angular resolution observations this can now be done as a function of radial
coordinate in the disk
\citep[e.g.,][]{Guilloteau:2011ek,Isella:2010ji,Perez:2012ii}, and shows that
bright rings have shallower slopes than the dark annuli between them
\citep[e.g.][]{ALMAPartnership:2015cg,Tsukagoshi:2016dl,Huang:2018fl}. However,
changes in the spectral slope can also be caused by optical depth effects. The
results of the ALMA Large Program DSHARP (\paperdsharpandrewst) show that these
dust rings often have an optical depth close to unity (\paperdsharphuangringst,
\paperdsharpdullemondt, \paperdsharpisellat).

With the detailed spatial information of the substructures in millimeter
continuum and line maps of protoplanetary disks that ALMA is providing, it
becomes increasingly important to discuss the details of the opacities used and
the methods applied to translate the observations into information about the
underlying dust grains.

The present paper is meant to give an overview of the methods used and choices
made by the DSHARP collaboration to make this translation. We do not claim in
any way that our choices and methods are better than those used by others, nor
that we can resolve any of the uncertainties of the opacities. Instead, this
paper describes our methods and choices, and we provide an easy-to-use python
module and a set of example calculations for the reproduction of these opacities
and variants of them, as well as for handling some of the optical depth effects
discussed in this paper.

The structure of this paper is as follows: In \autoref{sec:opacities} we discuss
our choices for computing the dust opacities, and present our Python tools that
are publicly available. In \autoref{sec:grain-size-aver} we apply these
opacities to simple size distribution models, starting with a standard power-law
model, and ending in \autoref{sec:meanopacs} with the analytic steady-state dust
coagulation/fragmentation size distribution of \citet[henceforth
\citetalias{Birnstiel:2011ks}]{Birnstiel:2011ks}, again including the
corresponding Python scripts. Finally, in \autoref{sec:layer-model} we present a
very simple model to link the observed thermal emission to the observed
extinction of back-side CO line emission (\paperdsharpisellat).

\begin{figure}[t]
 \includegraphics[width=\linewidth]{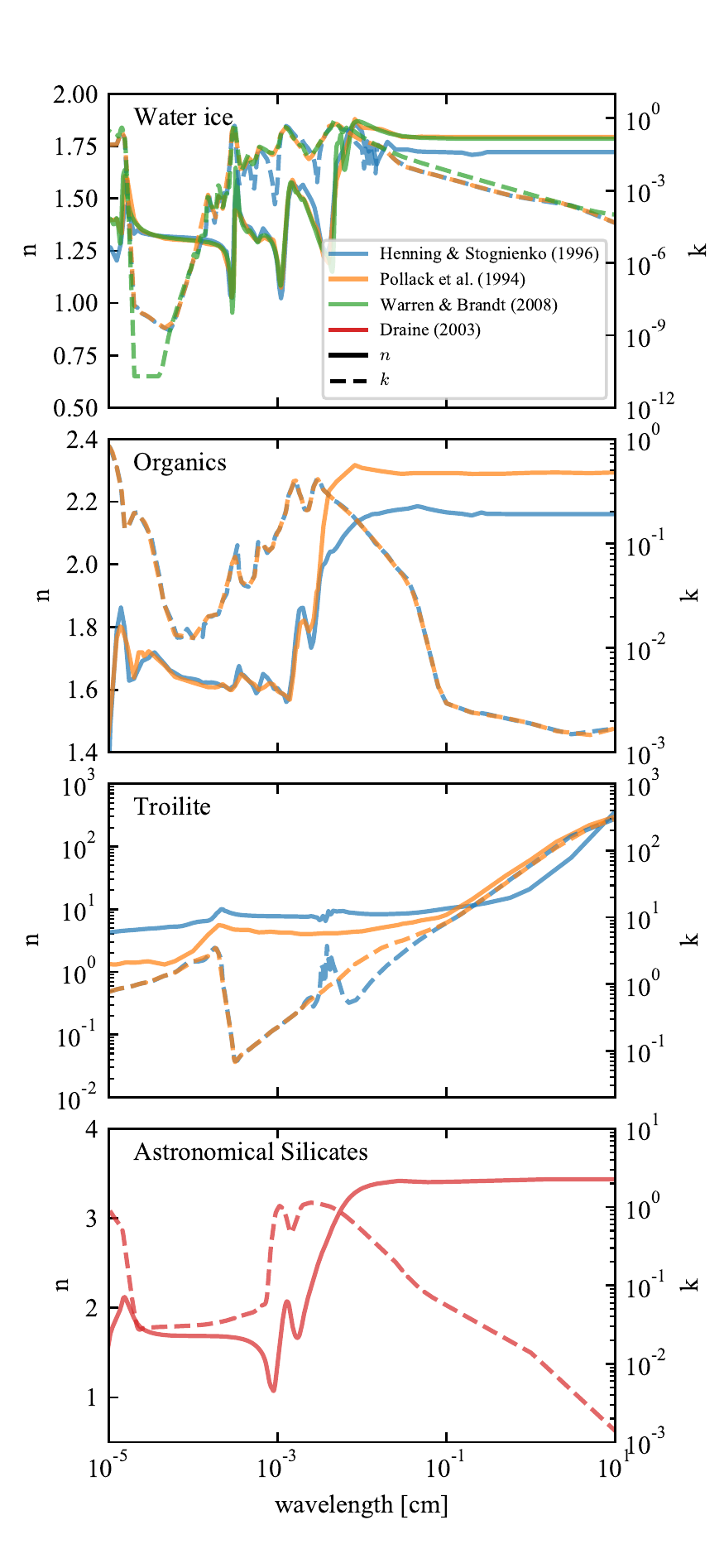}
 \caption{Optical constants used in this work (see \autoref{tab:composition}),
  compared to other literature data. Solid lines denote $n$ on the left axis and
  dashed lines denote $k$ on the right axis.\label{fig:opt_const}}
\end{figure}

\section{DSHARP Dust Opacities}
\label{sec:opacities}
As discussed above, the goal of this paper is not to provide a ``better'' dust
opacity model, but instead a transparent model based on open-source software
that is easy to reproduce or to modify. To this end, we follow seminal works
regarding protoplanetary disk composition and grain structure
\citep[][henceforth~\citetalias{Pollack:1994fn}]{Pollack:1994fn} that are widely
used  throughout the literature and we use updated optical constants where
available. To stay comparable to previously used opacities (and thus the
resulting mass or surface density estimates), we chose to assume particles
without porosity. This is a pragmatic choice instead of a realistic one for
protoplanetary disks since since at least the initial growth phase involves
larger porosities
\citep{Kempf:1999ie,Ormel:2007bh,Zsom:2010hg,Okuzumi:2012kd,Krijt:2015bu}.

\citetalias{Pollack:1994fn} and subsequent work by \citet{DAlessio:2001fk}
chose a mixture of water ice, astronomical silicates, troilite, and refractory
organic material. The water fraction that was used in those works (around 60\%
by volume) was, however in disagreement with typically observed disk spectral
energy distributions (SEDs), as pointed out by \citet{DAlessio:2006io} and
\citet{Espaillat:2010kc} who reduced the water fraction to 10\% of the value
used in \citetalias{Pollack:1994fn}. Since comets are thought to be a relatively
pristine sample of the planet forming material, we chose a water
fraction of 20\% by mass, in agreement with measurements of comet
67P/Churyumov-Gerasimenko \citep{Patzold:2016iv}.

To calculate mass absorption or scattering coefficients \kabs or \ksca, we
assume vacuum as embedding medium and furthermore need the complex refractive
indices $m(\lambda) = n(\lambda) + \mathrm{i}\,k(\lambda)$ which are functions
of wavelength $\lambda$. These refractive index data will be called
\textit{optical constants} for simplicity.

The optical constants of water used in \citet{DAlessio:2001fk} were from
\citet{Warren:1984ci} who gives tables for various temperatures.
\citet{Henning:1996ur} used optical constants from \citet{Hudgins:1993ks}
(amorphous ice at at 100 K) between 2.5 and \SI{200}{\mu m} and the constants
from \citetalias{Pollack:1994fn} for the remaining wavelengths ranges. In this
work, we use the more recent data from \citet{Warren:2008dn}. However
the differences to previous works are small (see \autoref{fig:opt_const}).

The astronomical silicates used in \citet{DAlessio:2001fk} took a constant $k$
value for  $\lambda>\SI{800}{\mu m}$. \citet{Henning:1996ur} argued (their
section 5.1) that $k\propto\lambda^{-1}$ is usually assumed, but that
\citet{Campbell:1969de} indeed measured a high value of $k = 0.05$ at
\SI{2.7}{mm} (see \autoref{fig:opt_const}). Nevertheless, we use the
opacities from \citet{Draine:2003di} for astronomical silicates without
increasing $k$.

For troilite and refractory organics, we use the optical constants from
\citet{Henning:1996ur}. For troilite, the constants are partly based on
\citet{Begemann:1994jc} (in the range of 10 to \SI{500}{\mu m}), with longer and
shorter ranges taken from \citetalias{Pollack:1994fn}. \citet{Henning:1996ur}
also performed Kramers-Kronig analysis on the organics optical constants of
\citetalias{Pollack:1994fn} which yielded little differences. The
\citet{Henning:1996ur} data sets for troilite and refractory organics are
available online, and are included in our opacity module with kind permission
from Thomas Henning. The optical constants used in this work are shown in
\autoref{fig:opt_const}.

Deriving optical constants for a mix of materials is a challenging task as it
depends on the detailed structure of the composite particle. No general solution
can be given. For more complex setups, computationally expensive numerical
models need to be used. For some limiting cases analytical expressions can be
given. These are typically called the Maxwell-Garnett rule (valid for inclusions
in a background ``matrix'') and the Bruggemann rule, for a homogeneous mix
without a dominant matrix. Details can be found in \citet{Bohren:1998wi} whose
notation we will follow.

If $f_i$ denotes the volume fractions of the $N$ inclusions ($i = 1\ldots N$) and
$\epsilon_i = m_i^2$ the dielectric functions of the inclusions\footnote{The
 dielectric functions of the inclusions $\epsilon_i$, the matrix $\epsilon_m$, or
 the mix $\bar\epsilon$ should not be confused with the absorption probability
 $\epsilon_\nu$ or
 $\epsilon_\nu^\mathrm{eff}$ used in later sections of this paper.} (with refractive indices $m_i$), while $f_m$ and
$\epsilon_m$ are the corresponding values for the matrix, then the
Maxwell-Garnett rule for spherical inclusions yields the mixed dielectric
function
\begin{align}
 \bar \epsilon & = \frac{(1 - f) \,  \epsilon_m + \sum_{i=1}^N{f_i \, \beta_i \, \epsilon_i}}
 {1 - f + \sum_{i=1}^N{f_i \, \beta_i}},
 \label{eq:maxwellgarnett}
\end{align}
where
\begin{align}
 \beta_i & = \frac{3 \, \epsilon_m }{\epsilon_i + 2 \, \epsilon_m}, \\
 f       & = \sum_{i=1}^N{f_i}.
\end{align}
For the case of the Bruggeman rule, the mixed material itself acts as
matrix and the mixed dielectric function can be calculated by solving for
$\bar\epsilon$ in the relation
\begin{align}
 \sum_{i=1}^N{f_i \, \frac{\epsilon_i-\bar \epsilon}{\epsilon_i + 2 \, \bar\epsilon}} = 0.
 \label{eq:bruggeman}
\end{align}

\begin{figure}
 \includegraphics[width=\linewidth]{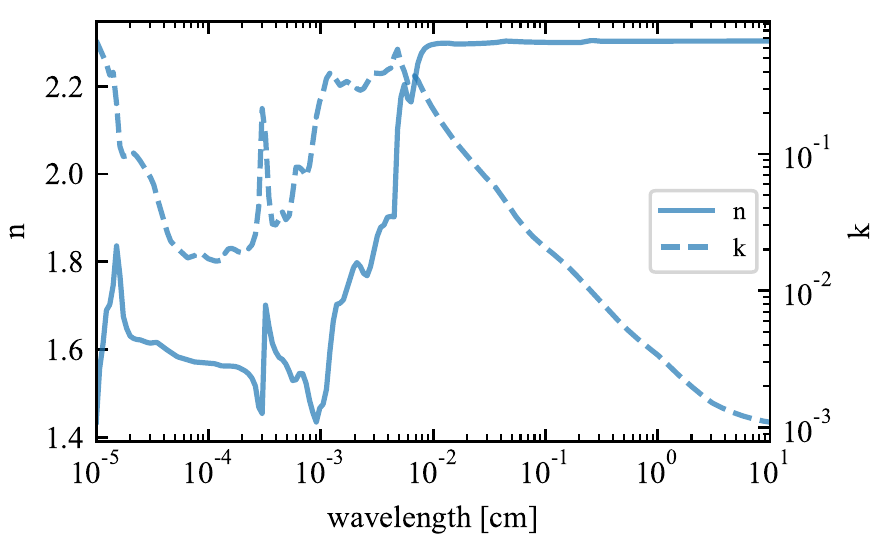}
 \caption{Effective medium optical constants that are used within the DSHARP collaboration,
  derived with the Bruggeman rule (\autoref{eq:bruggeman}).\label{fig:mix}}
\end{figure}

Both the Bruggeman and the Maxwell-Garnett rule are implemented in our opacity
module. However for the compact mixture of materials specified above and in
\autoref{tab:composition}, the Bruggeman rule is the appropriate choice. For this
dust model, the resulting effective medium optical constants are shown in
\autoref{fig:mix}.

\begin{figure}[bt]
 \includegraphics[width=\linewidth]{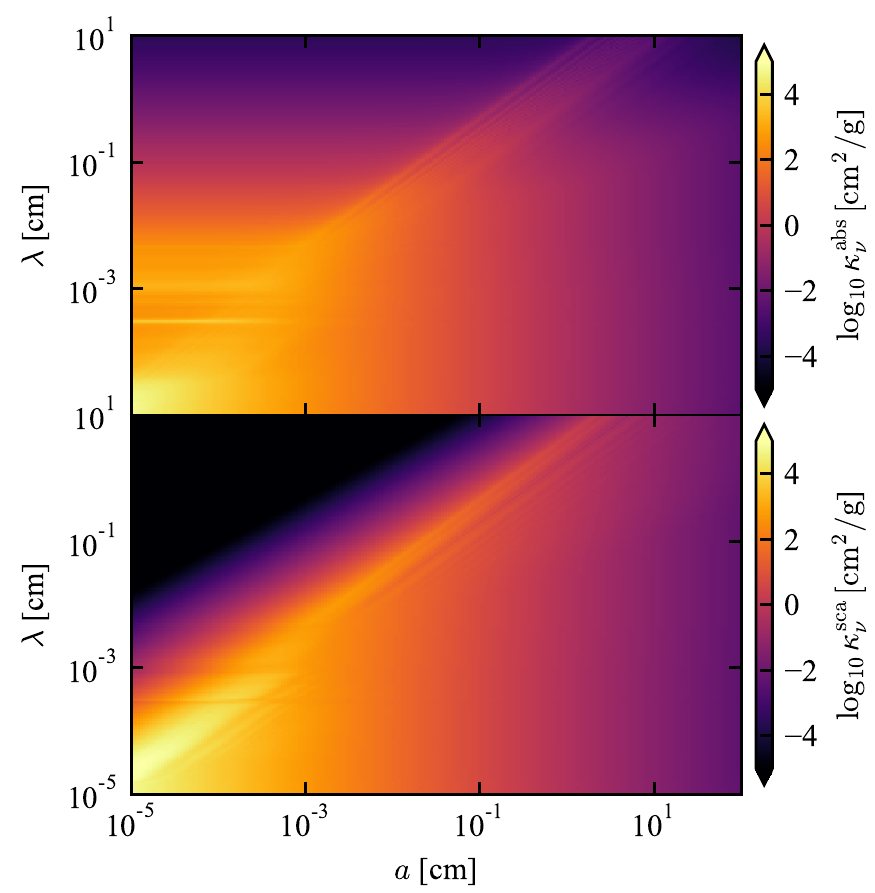}
 \caption{Absorption (top) and scattering (bottom) opacity as function of
  wavelength $\lambda$ and particle size $a$, based on Mie calculations using the
  optical constants from \autoref{fig:mix}.\label{fig:opacities}}
\end{figure}

Our opacity module includes a subroutine to do Mie opacity calculations using a
\texttt{Fortran90} subroutine for performance. This \texttt{Fortran90} code is
based on a \texttt{Fortran70} version by Bruce
Draine\footnote{\url{ftp://ftp.astro.princeton.edu/draine/scat/bhmie/bhmie.f}}
which itself is derived from the original Mie code published by
\citet{Bohren:1998wi}. To avoid strong and artificial Mie interferences, we do
not use single-grain-size opacities, but instead calculate the opacity for 40
linearly spaced bins within each grain size bin and average over those opacity
values to calculate an averaged opacity value for every size bin (each bin is
\SI{0.035}{dex} in size). The resulting absorption and scattering opacities are
shown in \autoref{fig:opacities} and are available from the module
repository\footnote{\url{https://github.com/birnstiel/dsharp_opac}}.

\begin{deluxetable*}{lllll}
 \tablecaption{Dust composition used in the DSHARP collaboration\label{tab:composition}}
 \tablehead{
  \colhead{Material} & \colhead{References} & \colhead{bulk density} &
  \colhead{mass fraction} & \colhead{vol. fraction}\\
  & & [g/cm$^3$] \\
 }
 \startdata
 Water ice               & \citet{Warren:2008dn}  & 0.92 &  0.2000 & 0.3642 \\
 Astronomical Silicates  & \citet{Draine:2003di}  & 3.30 &  0.3291 & 0.1670 \\
 Troilite                & \citet{Henning:1996ur} & 4.83 &  0.0743 & 0.0258 \\
 Refractory organics     & \citet{Henning:1996ur} & 1.50 &  0.3966 & 0.4430 \\
 \enddata
 \tablecomments{The bulk density of the mix is $\rhos = \SI{1.675}{g.cm^{-3}}$.}
\end{deluxetable*}

\section{Grain-size averaged opacities}
\label{sec:grain-size-aver}
It is known that dust grains in protoplanetary disks are not ``mono-disperse'',
i.e., at a given radius in the disk the dust does not consist of only a single
size, or a narrow size distribution. Perhaps the most spectacular evidence of
this is found in the source IM Lup. When observed at near-infrared wavelengths,
this disk shows a strongly flaring geometry \citep{Avenhaus:2018et}. Clearly, a
substantial amount of fine-grained dust is suspended several scale heights above
the mid-plane, and is continuously replenished by turbulent stirring. When
observed at millimeter wavelengths, however, we see a disk with small-scale ring
and spiral substructure that can only be explained if the geometry of this dust
layer is vertically geometrically thin
\citep[e.g.,][\paperdsharphuangspiralst]{Pinte:2008gi}. This must be of a grain
population that is vastly larger (and/or more compact) than the grains seen in
the near-infrared. Posed more precisely: IM Lup features at least two dust grain
populations, one with very small Stokes number, and therefore vertically
extended, and one with much larger Stokes number, and therefore vertically flat
due to settling.

It is reasonable to expect that the dust population in fact consists of a
continuous size distribution instead of just two distinct sizes. This is what is
expected from models of dust coagulation which include fragmentation
\citep{Weidenschilling:1984ih,Dullemond:2005hf,Brauer:2008bd,Birnstiel:2010eq}.
These populations change with time, as the grains drift and grow at different
rates. The complexity of this process makes it hard to define a simple
``one-size-fits-all'' dust opacity model to be used for interpreting millimeter
continuum maps of protoplanetary disks. On the other hand, detailed
coagulation/fragmentation modeling is numerically expensive, and it is not
feasible to analyze all data with such full-fledged models.

Many authors use therefore a compromise by assuming that the dust grain size
distribution follows a simple power-law with a cut-off at small and large grain
sizes,
\begin{align}
 n(a) \propto \, \begin{cases}
 a^{-q} & \text{for} \, a_\mathrm{min} \leq a \leq a_\mathrm{max} \\
 0      & \text{else},
 \end{cases}
 \label{eq:powerlaw}
\end{align}
where the total dust density is defined as $\rho_\mathrm{d} =
\int_0^\infty\,n(a)\,m(a)\,\mathrm{d}a$ with $m(a)$ being the mass of a dust
particle of radius $a$. The resulting opacity at (sub-)millimeter wavelengths
is found to be less sensitive to the minimum grain radius \amin, but much more
so to the maximum particle size \amax as well as the power-law index $q$
\citep{Ricci:2010gc,Draine:2006is}.

\begin{deluxetable*}{lp{4cm}p{4cm}}
 \tablecaption{Dust size distributions used throughout this paper.\label{tab:distributions}}
 \tablehead{
  \colhead{Acronym} & \colhead{Description} & \colhead{References}
 }
 \startdata
 MRN  & power-law, $q=3.5$ or                                             & \citet{Mathis:1977hp},\\
 ~    & $q=2.5$                                                           & \citet{DAlessio:2001fk},\\
 ~    & ~                                                                 & \autoref{eq:powerlaw}\\
 B11  & analytic fits to detailed simulations of growth and fragmentation & \citet{Birnstiel:2011ks}\\
 B11S & simplified versions of B11                                        & \autoref{sec:appendixA}
 \enddata
\end{deluxetable*}

The index $q$ was found to be around $3.5$ for interstellar extinction
measurements \citep[e.g.,][henceforth~\citetalias{Mathis:1977hp}]{Mathis:1977hp}
which is consistent with collisional cascades
\citep{Dohnanyi:1969dw,Tanaka:1996ip} and also found to be consistent with
sub-millimeter observations of debris disks \citep{Ricci:2015ce}. However, the
physics of debris disks is very different from gaseous protoplanetary disks. One
way out is to use simplified dust coagulation/fragmentation models.  For
instance, \citetalias{Birnstiel:2011ks} presented an analytic multi-power-law
fit to the results of the full-fledged numerical dust coagulation/fragmentation
models. A summary of the size distributions and the acronyms used throughout
this paper can be found in \autoref{tab:distributions}.

In the following, we will use the DSHARP opacity model of
\autoref{sec:opacities} and apply it to a simple power-law size distribution. We
compare those results to the ones obtained if the analytic coagulation model
fits of \citetalias{Birnstiel:2011ks} are used. The Python script for creating
the resulting size-averaged opacities is publicly available in the module
repository.

The total absorption opacity \kabst of a particle size
distribution $n(a)$ at frequency $\nu$ is calculated from the size-dependent
opacity $\kabs(a)$ via

\begin{align}
 \kabst = \frac
 {\int_{\amin}^{\amax} n(a) \, m(a) \, \kabs(a) \, \mathrm{d}a}
 {\int_{\amin}^{\amax} n(a) \, m(a) \, \mathrm{d}a}.
\end{align}

\begin{figure}
 \includegraphics[width=\linewidth]{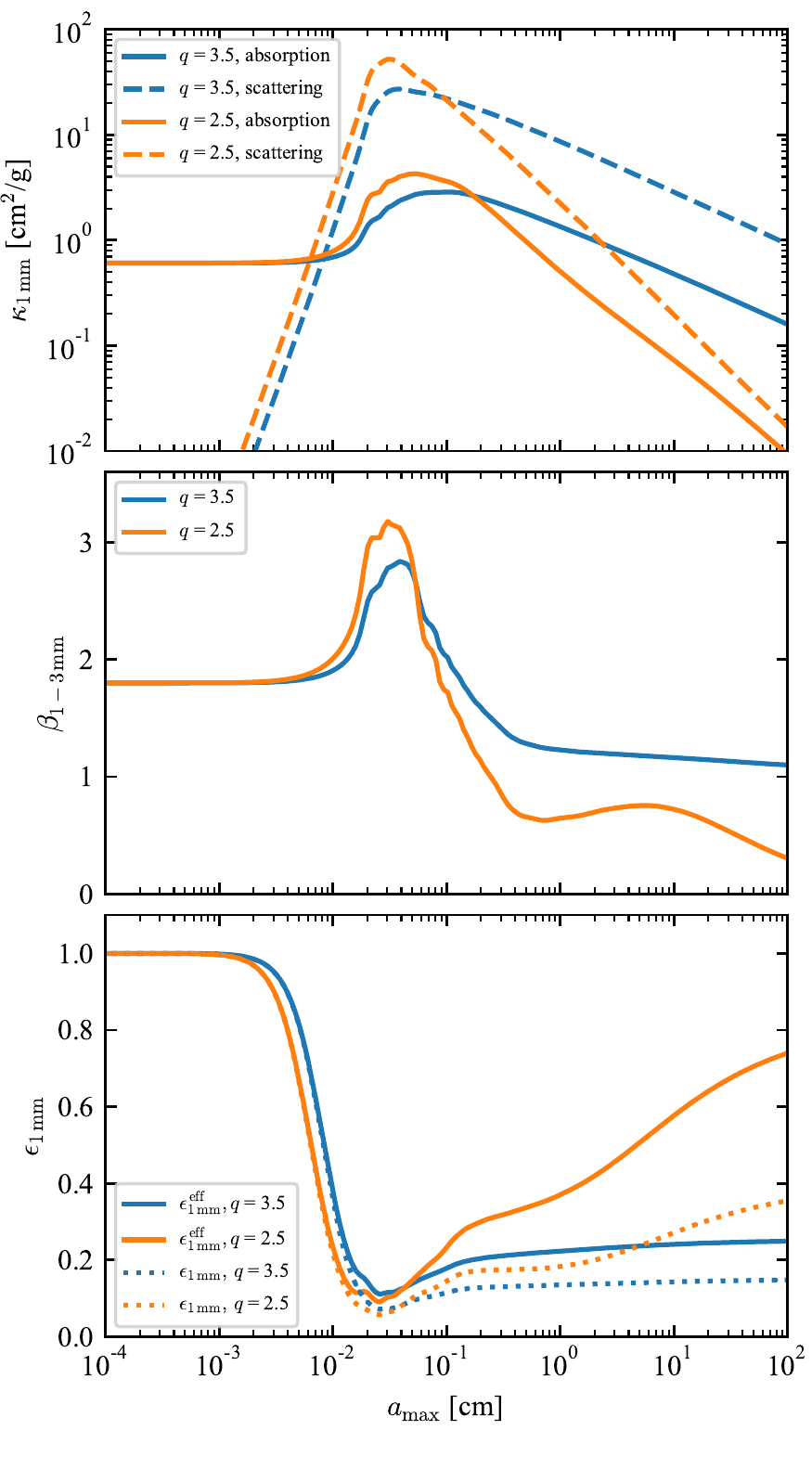}
 \caption{Particle size averaged opacities. Top: scattering (\kscat) and
  absorption (\kabst) opacity at \SI{1}{mm}. Middle: spectral index $\beta$
  measured at 1-\SI{3}{mm}. Bottom: extinction probability
  $\epsilon^\mathrm{eff}_{\SI{1}{mm}}$ (see \autoref{sec:layer-model}). The
  assumed size distribution for these averaged properties follows a power-law
  $n(a) \propto a^{-q}$ from the minimum size of \SI{e-5}{cm} up to maximum size
  $a_\mathrm{max}$. Blue lines denote the MRN-slope of $q=3.5$, orange lines
  correspond to $q=2.5$.\label{fig:size_average}}
\end{figure}

\begin{figure}
 \includegraphics[width=\linewidth]{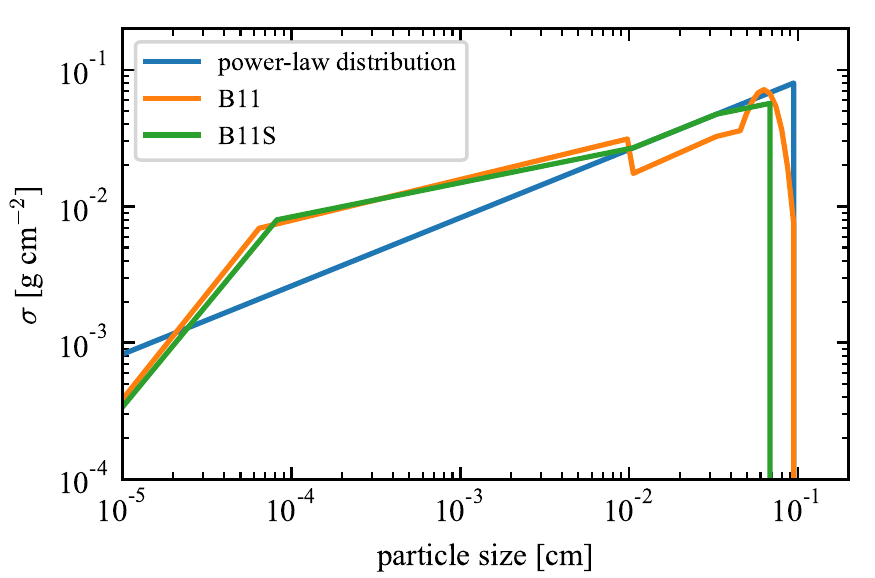}
 \caption{Particle size distributions used in \autoref{sec:grain-size-aver}.
  $\sigma(a)$ is the surface density per logarithm in particle size, see
  \autoref{eq:sigma-def}. The blue line is a truncated power-law with $q=3.5$.
  The orange line is a size distribution fit in coagulation/fragmentation
  equilibrium from \citetalias{Birnstiel:2011ks}, where parameters were chosen to
  result in a maximum particle size of \SI{1}{mm}. The green line labeled B11S is
  a simplified version of the \citetalias{Birnstiel:2011ks} fit using only a
  broken power-law. This neglects finer details of the fit, but avoids
  calculating collision velocities.\label{fig:size_distri}}
\end{figure}

The top panel in \autoref{fig:size_average} shows the total absorption and
scattering opacities at a wavelength of \SI{1}{mm} for a particle size
distribution with $\amin = \SI{e-5}{cm}$ as function of \amax. The bottom panel
shows the spectral index $\beta = \partial \ln \kabst / \partial \ln \nu$.
Similar trends as in \citet{Ricci:2010gc} are observed: changes in the size
distribution index $q$ mainly affect the asymptotic behavior at long wavelengths
and the strength of the Mie interference at $\amax \sim \frac{\lambda}{2 \pi}$.
\autoref{fig:size_average} also shows that for size distributions that extend up
to $\amax \gtrsim \SI{100}{\mu m}$, the scattering opacity \kscat exceeds the
absorption opacity \kabst.

Three different particle size distributions were chosen, that have the same
maximum particle size \amax of \SI{1}{mm}, however one follows the MRN-like
size-exponent of $q=3.5$, while the other two distribution are steady-state size
distributions where continuous particle growth and fragmentation lead to a
stationary size distribution. The first of these (orange line in
\autoref{fig:size_distri}) is from detailed analytical fits to numerical
simulations from  \citetalias{Birnstiel:2011ks}. The second steady-state
distribution, shown in green in \autoref{fig:size_distri}, is a simplified
version of these fits. This implements only the piecewise power-laws from
\citetalias{Birnstiel:2011ks} and ignores finer details. This avoids calculating
collision velocities for all particle sizes and thus makes the calculation
easier and faster (see \autoref{sec:appendixA}). This simplified fit still
captures the important aspects of the simulated distributions much better than the
two-power-law fits used in \citet{Birnstiel:2015cl} and \citet{Ormel:2013ha}.
Especially for large particle sizes, the two-power-law fits can underpredict the
number of small particles available.

\begin{figure*}[t]
 \includegraphics[width=\linewidth]{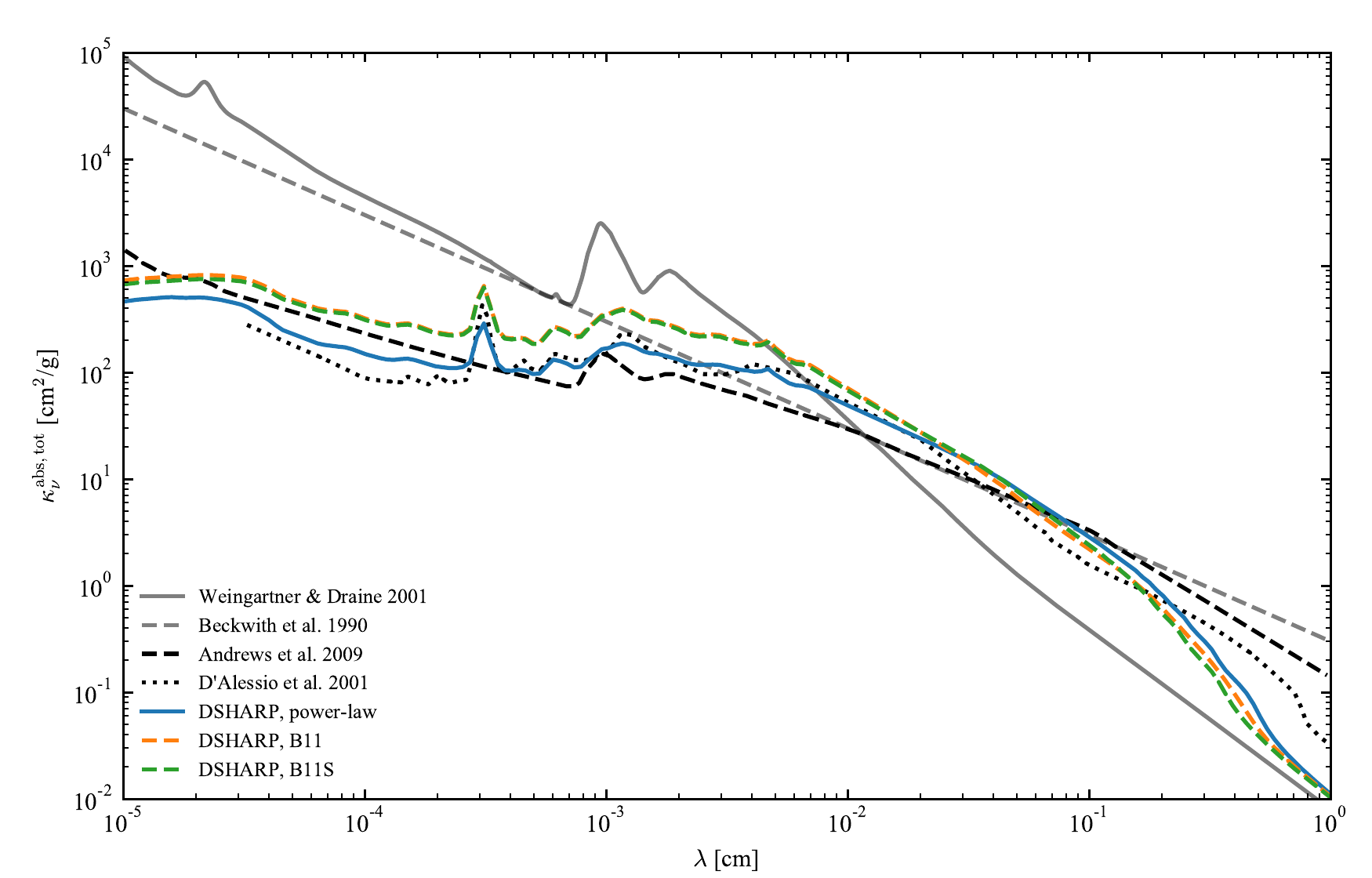}
 \caption{Wavelength dependency of size averaged absorption opacities using a
  power-law size distribution ($\amin=\SI{e-5}{cm}$, $\amax=\SI{1}{mm}$, $q=3.5$)
  and opacity values used in the literature for
  comparison.\label{fig:comparison}}
\end{figure*}

Wavelength-dependent opacities for all three distributions have been calculated
and are shown in \autoref{fig:comparison} in comparison with opacities used in
the literature. It can be seen that the overall behavior is -- by construction
-- similar to the opacities in \citet{DAlessio:2001fk} or \citet{Andrews:2009jo}
with slightly different behavior at long wavelengths. Differences in the $\mu$m
wavelength range are mainly due to the different amounts of small grains present
in the distributions due to the knee in the steady state distributions. This
comes from the fact that smaller particles that move at higher Brownian motion
velocities are more efficiently incorporated into larger particles. It can be
seen that the two different fitting methods (labeled B11 and B11S) yield
virtually identical opacities. The small differences to the simple MRN-power-law
stems from the fact that parameters were chosen to yield the same \amax. As seen
from \autoref{fig:size_average}, \amax is the most important parameter
influencing the size-averaged opacity.

\section{Mean Opacities of steady-state size distributions}
\label{sec:meanopacs}

\begin{figure}[bt]
 \includegraphics[width=\linewidth]{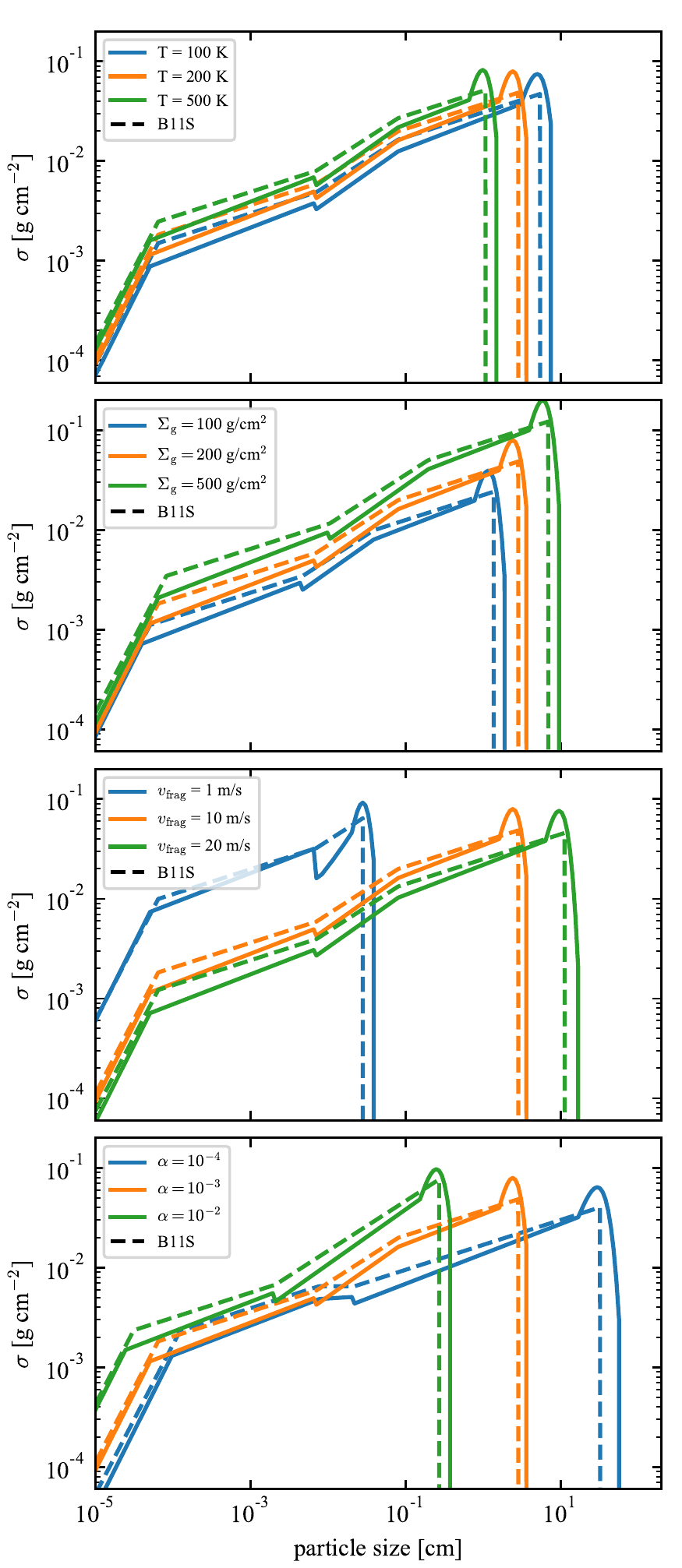}
 \caption{Comparison of fitting functions of \citetalias{Birnstiel:2011ks} and the simplified fitting function (see \autoref{sec:appendixA}). The fiducial model is denoted by the orange line and its parameters are given in \autoref{sec:meanopacs}.\label{fig:distri_panels}}
\end{figure}

The simplified fits labeled B11S are compared to the more detailed fits from
\citetalias{Birnstiel:2011ks} in \autoref{fig:distri_panels}. Given the
uncertainty in the details of collision models
\citep[see,][for~example]{Guttler:2010ia,Windmark:2012gi}, and for ease of
reproduction, we will be using the B11S fits in the following. They are
explained in \autoref{sec:appendixA}. Throughout this paper, we assume a
dust-to-gas mass ratio of 0.01 and consider size distributions integrated over
height; settling will cause the size distribution to depend on the vertical
position above the mid-plane.

\autoref{fig:distri_panels} shows how the particle size distribution in
steady-state varies with the gas temperature $T$, the gas surface density
$\Sigma_\mathrm{g}$, the fragmentation threshold velocity $v_\mathrm{frag}$, and
the turbulence parameter $\alpha$ \citep{Shakura:1973uy}. It can be seen, that
the position in the knee of the distribution at size $a_\mathrm{BT}$ (cf.
Equation 37 in \citetalias{Birnstiel:2011ks}) has only a weak dependence on
those parameters, however that the maximum particle size \amax is a strong
function of these parameters (quadratic in $v_\mathrm{f}$, linear in all
others). As such, it also affects the Planck and Rosseland mean opacities,
\begin{align}
 \bar \kappa_\mathrm{P}(T) & = \frac
 {\int_0^\infty \kabst\, B_\nu(T)\,\mathrm{d}\nu}
 {\int_0^\infty  B_\nu(T)\,\mathrm{d}\nu},\\
 \bar \kappa_\mathrm{R}(T) & = \left( \frac
 {\int_0^\infty \frac{1}{\kextt} \, \frac{\mathrm{d}B_\nu(T)}{\mathrm{d}T}\,\mathrm{d}\nu}
 {\int_0^\infty  \frac{\mathrm{d}B_\nu(T)}{\mathrm{d}T}\,\mathrm{d}\nu}
 \right)^{-1},
\end{align}
since now not only the Planck spectrum $B_\mathrm{\nu}(T)$ is temperature
dependent, but also the size-averaged opacities \kabst, and \kextt.
This means that the mean opacities are additionally dependent on other physical parameters, $\Sigma_\mathrm{g}, v_\mathrm{frag}$, and $\alpha$.

\begin{figure}[t]
 \includegraphics[width=\linewidth]{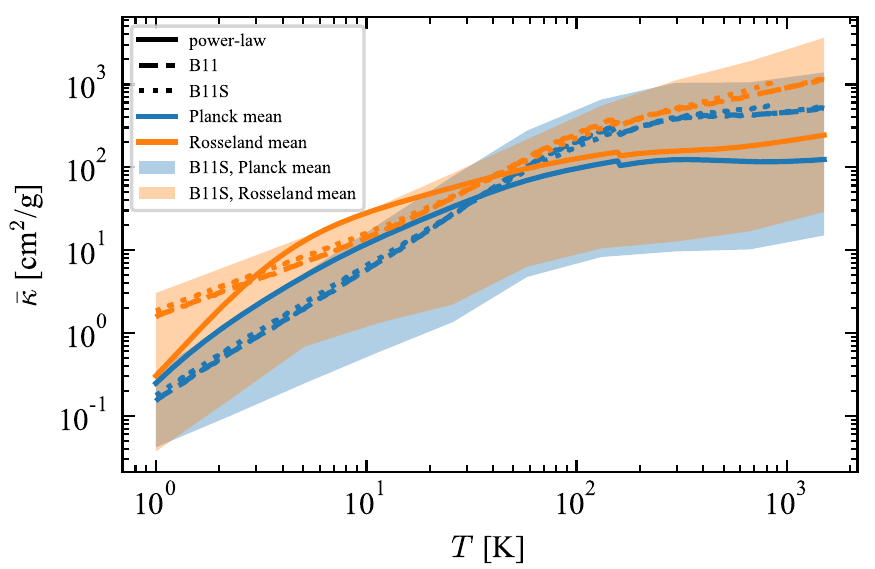}
 \caption{Planck (blue) and Rosseland (orange) mean opacities for steady state
  size distributions (dashed and dotted lines) and for a power-law distribution
  with fixed $\amax=\SI{1}{mm}$ (solid lines). The fits (dashed and dotted lines)
  used fixed parameters as in \autoref{fig:size_distri}, only varying the
  temperature. The shaded regions show the range of opacities if the other
  parameters affecting the fits are varied within reasonable
  ranges (see \autoref{sec:meanopacs}). For temperatures above the water sublimation temperature, the water ice was removed from the material mix.\label{fig:mean_opacities}}
\end{figure}

As an example, \autoref{fig:mean_opacities} shows the Rosseland and Planck mean
opacities for two cases: a \citetalias{Mathis:1977hp}-size distribution (as in
\autoref{fig:size_distri}) and the steady-state distributions
\citetalias{Birnstiel:2011ks} and B11S as a function of temperature. It can also
be seen, that differences between the two steady-state distributions are small,
allowing the simpler model to be used without caveats. It can be seen that for
the fiducial  values of $M_\star = M_\odot$, $r = \SI{1}{au}$, $v_\mathrm{frag}
= \SI{100}{cm/s}$, and $\alpha = \SI{e-3}{}$, high-temperature mean opacities
are generally higher for steady-state distributions owing to the fact that the
knee at $\mu$m sizes produces more small grains for the same \amax than a single
power-law size distribution. Furthermore high temperatures tend to produce
smaller \amax than \SI{1}{mm} which additionally increases the amount of small
particles that contribute most to the mean opacities. The shaded areas in
\autoref{fig:mean_opacities} are covering the ranges $\SI{1}{K} \leq T \leq
\SI{1500}{K}$, $\SI{50}{cm.s^{-1}} \leq v_\mathrm{frag} \leq
\SI{3e3}{cm.s^{-1}}$, $\SI{e-5}{} \leq \alpha \leq \SI{e-1}{}$,
$\SI{1}{g.cm^{-2}} \leq \Sigma_\mathrm{g} \leq \SI{e4}{g.cm^{-2}}$. It should be
noted that the distributions discussed here only apply to those parts of the
disk where particles reach the fragmentation barrier $a_\mathrm{frag}$ -- this is
only possible if 1) collision velocities are high enough (i.e. the root in
\cref{eq:fragbarrier2} is real), and 2) fragmentation  is more important in
limiting particle growth than drift, which is the case if
\citep[see][]{Birnstiel:2015cl}
\begin{equation}
 \alpha > \frac{\left|\gamma\right|\,\Sigma_\mathrm{g}}{3\,\Sigma_\mathrm{d}} \left(\frac{v_\mathrm{frag}}{V_\mathrm{K}}\right),
\end{equation}
where $\gamma$ is the radial logarithmic pressure gradient $\partial \ln
P/\partial \ln r$ and $V_\mathrm{K}$ the Keplerian orbital velocity. If the
drift limit applies, the size distribution will contain more mass at the largest
sizes and is more strongly dependent on global redistribution of particles.
These non-local processes can be approximated as in \citet{Birnstiel:2015cl},
however the accuracy of these approximations are likely only good enough for
applying them to the long-wavelength opacity. Short wavelengths are too
sensitive to small changes in the amount of small grains, which in turn depend
sensitively on radial mixing, and details of the collisional model.

\section{Dust emission and extinction from a thin dust layer with scattering}
\label{sec:layer-model}

For protoplanetary disks, it is mostly assumed that only the absorption opacity
\kabs, not the scattering opacity, matters. For optically thin dust layers this
is indeed appropriate. Recently, the importance of scattering and its effects on
(sub-)millimeter polarization of disks was pointed out by
\citet{Kataoka:2015hl}. In the DSHARP campaign we have seen that the optical
depths are not that low ($0.1 \lesssim \tau \lesssim 0.6$, see
\paperdsharpdullemondt). Furthermore, the CO line extinction data of HD 163296
discussed by \paperdsharpisella suggest that the dust layer has an extinction
optical depth close to unity (see also \paperdsharpguzmant). Even if the
absorption optical depth is substantially below 1, the total extinction
(absorption + scattering) can easily exceed unity, if the grains are of similar
size to the wavelength. For $a\simeq \lambda/2\pi=\SI{0.13}{cm}/2\pi =
\SI{0.02}{cm}$ the albedo of the grain can, in fact, be as high as 0.9 (see
\autoref{fig:size_average} and \autoref{sec:appendixB}).

The inclusion of scattering complicates the radiative transfer equation
enormously. Strictly speaking a full radiative transfer calculation, for
instance with a Monte Carlo code, is necessary. However, in the spirit of this
paper we wish to find a simple approximation to handle this without resorting to
complex numerical simulation.

There are two issues to be solved: One is: what thermal emission will a
non-optically-thin dust layer produce if scattering is taken into account? The
other is: how do we compute the extinction coefficient to be used in the CO line
extinction analysis?

For the first issue, we will outline here a simple two-stream radiative transfer approach to the
problem. We will assume that the dust seen in the ALMA observations is located
in a geometrically thin layer at the mid-plane, so we can use the 1-D slab
geometry approach. We will assume that the scattering is isotropic. This may be
a bad approximation, especially for $2\pi a\gg \lambda$. To reduce the impact of
this approximation we replace the scattering opacity $\ksca$
with
\begin{equation}
 \kscaeff = (1-g_\nu)\, \ksca
\end{equation}
where $g_\nu$ is the usual forward-scattering parameter (the expectation value
of $\cos\theta$, where $\theta$ is the scattering angle). According to
\citet{Ishimaru:1978hz}, this approximation works well in optically thick media.

\begin{figure}[tb]
 \centerline{\includegraphics[width=0.47\textwidth]{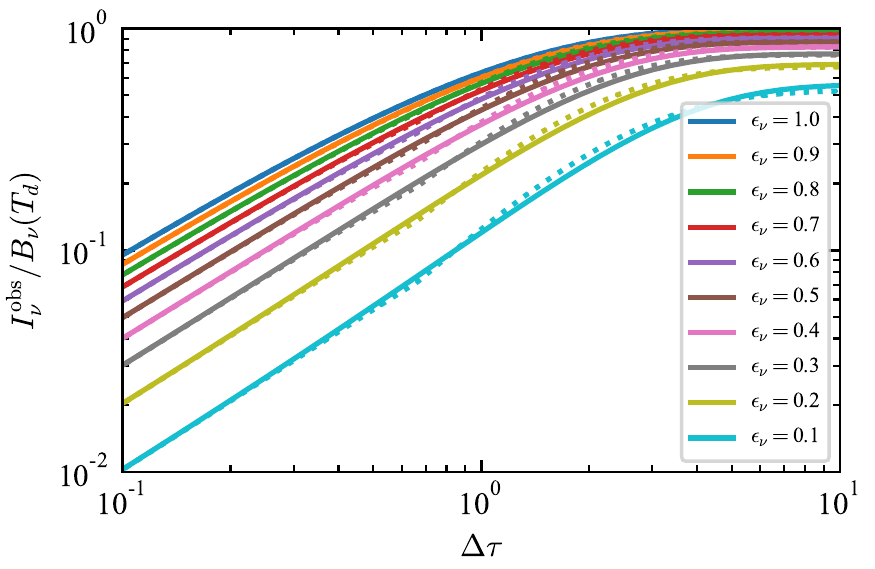}}
 \caption{\label{fig-scatter-parscan}The intensity $I_\nu$, in units of the
  Planck function, emerging from a slab seen face on, with total optical depth
  $\Delta\tau$, a constant temperature, and an albedo of
  $\eta_\nu=1-\epsilon^\mathrm{eff}_\nu$. See \autoref{sec:layer-model} for
  details. The solid lines are the results of numerical integration of
  \autoref{eq-fte-slab-with-scat}. The dotted lines are the result of the
  modified Eddington-Barbier approximation
  (\autoref{eq-modif-eddington-barbier}).}
\end{figure}

We will now follow the two-stream / moment method approach from
\citet{Rybicki:1991vm} to derive the solution to the
emission/absorption/scattering problem in this slab. The slab is put between
$z=-\tfrac{1}{2}\Delta z$ and $z=+\tfrac{1}{2}\Delta z$ and we assume a constant
density of dust between these two boundaries. The mean intensity $J_\nu(z)$ of
the radiation field then obeys the equation
\begin{equation}
 \frac{1}{3}\frac{d^2J_\nu}{d\tau_\nu^2} = \epsilon_\nu \big(J_\nu-B_\nu(T_d)\big)
\end{equation}
where
\begin{equation}
 \tau_\nu=\rho_d\,(\kabs+\kscaeff) z
 \equiv \rho_d\, \ktot z
\end{equation}
with $\rho_d$ being the dust density, and
\begin{equation}
 \epsilon^\mathrm{eff}_\nu = \frac{\kabs}{\kabs+\kscaeff}
\end{equation}
The boundary conditions at $z=\pm\tfrac{1}{2}\Delta z$ are
\begin{equation}
 \frac{dJ_\nu}{d\tau_\nu} = \mp \sqrt{3}J_\nu
\end{equation}
This leads to the following solution:
\begin{equation}
 \frac{J_\nu(\tau_\nu)}{B_\nu(T_d)} =  1-b\,
 \left(e^{-\sqrt{3\epsilon^\mathrm{eff}_\nu}\left(\tfrac{1}{2}\Delta\tau-\tau_\nu\right)}+e^{-\sqrt{3\epsilon^\mathrm{eff}_\nu}\left(\tfrac{1}{2}\Delta\tau+\tau_\nu\right)}\right)
\end{equation}
where $\Delta\tau = \rho_d\, \ktot \Delta z$, and $b$ is
\begin{equation}
 b = \left[(1-\sqrt{\epsilon^\mathrm{eff}_\nu})e^{-\sqrt{3\epsilon^\mathrm{eff}_\nu}\Delta\tau} + 1 + \sqrt{\epsilon^\mathrm{eff}_\nu}\right]^{-1}
\end{equation}
Given this solution for the mean intensity $J_\nu(\tau_\nu)$ we can now numerically integrate
the formal transfer equation along a single line of sight passing through the slab at an
angle $\theta$:
\begin{equation}\label{eq-fte-slab-with-scat}
 \mu\frac{dI_\nu(\tau_\nu)}{d\tau_\nu} = \epsilon^\mathrm{eff}_\nu B_\nu(T_d) + (1-\epsilon^\mathrm{eff}_\nu) J_\nu(\tau_\nu)
 -I_\nu(\tau_\nu)
\end{equation}
where $\mu=\cos\theta$. We start at $\tau_\nu=-\tfrac{1}{2}\Delta \tau$ with
$I_\nu=0$, and integrate to $\tau_\nu=+\tfrac{1}{2}\Delta \tau$. The resulting
$I_\nu^{\mathrm{out}}=I_\nu(\tfrac{1}{2}\Delta \tau)$ is the intensity that is
observed by the telescope. An approximation for $I_\nu^{\mathrm{out}}$ which
works well to within a few percent is the following modified version of the
Eddington-Barbier approximation:
\begin{equation}\label{eq-modif-eddington-barbier}
 I_\nu^{\mathrm{out}} \simeq \left(1-e^{-\Delta\tau/\mu}\right)\,S_\nu\left((\tfrac{1}{2}\Delta\tau-\tau_\nu)/\mu=2/3\right)
\end{equation}
where
\begin{equation}
 S_\nu(\tau_\nu) = \epsilon^\mathrm{eff}_\nu B_\nu(T_d) + (1-\epsilon^\mathrm{eff}_\nu) J_\nu(\tau_\nu)
\end{equation}
is the source function. In the optically thin case, when $\Delta\tau/\mu<2/3$, the value of $S_\nu$ is taken at the edge of the slab. The results are shown in \autoref{fig-scatter-parscan}.

For small optical depth ($\Delta\tau\ll 1$) the role of scattering vanishes,
and the solution approaches: $I_\nu^{\mathrm{out}} \rightarrow \epsilon^\mathrm{eff}_\nu\Delta\tau
B_\nu(T_d)/\mu$.  This is the same limiting solution as when
$\ksca$ is set to zero but $\kabs$ is
kept the same. For high optical depth the outcoming intensity does not saturate
to the Planck function, but a bit below, if the albedo is non-zero. This is the
well-known effect that scattering makes objects appear cooler than they really
are.

Now let us turn to the second issue to be considered: how to calculate
the actual extinction coefficient for the CO line extinction analysis.
At first sight the answer is simple:
\begin{equation}
 \kext = \kabs + \ksca
\end{equation}
where we use $\ksca$, not $\kscaeff$, for the scattering.  This is, in fact, the
correct answer for the case in which the emitting CO line emitting layer is far
behind the extincting dust layer, where ``far'' is defined in comparison to the
width of the extincting dust ring. If this condition is, however, not met, then
the scattering will not reduce the CO line intensity as much as naively
expected. CO line photons that are heading elsewhere might then, in fact, get
scattered into the line of sight. This effect is exacerbated for the case
of small-angle scattering. One can therefore argue that this effect
reduces the extinction of the CO line emission by the dust layer. For the
case of HD 163296 (\paperdsharpisellat{}) it appears that these effects
are not too strong, so a first analysis without accounting for this is
in order. A final answer may, however, require a full treatment of 3-D
radiative transfer.

\section{Summary}
\label{sec:summary}
In this paper we present methods for translating observed dust emission and
extinction used in the DSHARP campaign, which can also be used by other work. The
DSHARP opacities presented here are merely a choice, based on reasonable
assumptions. They provide the standard used within the DSHARP campaign.

We further explore how steady-state size distributions in a
coagulation-fragmentation equilibrium affect dust opacities by introducing
dependencies on temperature, surface density, turbulence and material
properties. We provide simplified fits to analytical functions and show the
range of Rosseland and Planck mean opacities covered for a wide range of
parameter choices.

Given the large albedo at (sub-)millimeter wavelength ranges, we derive
solutions to the radiative transfer equation for a homogeneous medium with
scattering and absorption. Together with the DSHARP dust model, and based on the
measurements of \paperdsharpisella, we find that the particles in the rings of
HD163296 should be at least \SI{0.2}{cm} of size.

Along with this paper, we present publicly available Python scripts that contain
the optical data of many literature materials. In addition to that, functions
are available for  mixing optical constants with effective medium theory, for
calculating opacities using Mie theory, and for averaging opacities over
particle size distributions. Implementations of the steady state particle size
distributions discussed in this paper are also included. The online material
includes these python modules, scripts for generating the  results and figures
of this paper, and the opacity tables. These materials will likely be extended
in the future but the version used in this paper is available at
\citet{github_dsharp_opac}. Additional material will be described in appendices
of future papers, as they become available.

\acknowledgments
We like to thank Ryo Tazaki, Akimasa Kataoka and Satoshi Okuzumi for helpful discussions and Thomas Henning for providing the optical constants data used in this work.
T.B. acknowledges funding from the European Research Council (ERC) under the European Union's Horizon 2020 research and innovation programme under grant agreement No~714769.
C.P.D. acknowledges support by the German Science Foundation (DFG) Research Unit FOR~2634, grants DU~414/22-1 and DU~414/23-1.
Z.Z. and S.Z. acknowledges support from the National Aeronautics and Space Administration through the Astrophysics Theory Program with Grant No.~NNX17AK40G and Sloan Research Fellowship. Simulations are carried out with the support from the Texas Advanced Computing Center (TACC) at The University of Texas at Austin through XSEDE grant TG- AST130002.
S. A. and J. H. acknowledge funding support from the National Aeronautics and Space Administration under grant No.~17-XRP17$\_$2-0012 issued through the Exoplanets Research Program.
J.H. acknowledges support from the National Science Foundation Graduate Research Fellowship under Grant No. DGE-1144152.
A.I. acknowledges support from the National Aeronautics and Space Administration under grant No.~NNX15AB06G issued through the Origins of Solar Systems program, and from the National Science Foundation under grant No.~AST-1715719.
L.P. acknowledges support from CONICYT project Basal AFB-170002 and from FCFM/U. de Chile Fondo de Instalaci\'on Acad\'emica.

\software{
 {\tt Numpy} \citep{numpy},
 {\tt Matplotlib} \citep{matplotlib},
 {\tt Astropy} \citep{astropy}.
}

\appendix

\section{Simplified Steady State Distributions}
\label{sec:appendixA}

The simplified version of the \citetalias{Birnstiel:2011ks} steady-state distributions used in this work are defined as a broken but continuous power-law as function of particle size,
\begin{align}
 \sigma(a) =
 \begin{cases}
 a^{p-1} & \text{for} \quad a\leq a_\mathrm{frag} \\
 0       & \text{else}
 \end{cases}
\end{align}
where the exponents $p$ are changing at specific sizes. They are chosen according to this algorithm
\begin{align}
 \begin{split}
   & \mathbf{if} (a < a_\mathrm{BT}):                          \\
   & \qquad p = \frac{3}{2} \quad \text{or} \quad \frac{5}{4}  \\
   & \mathbf{else if} (a \leq a_{12}):                         \\
   & \qquad p = \frac{1}{4}  \quad \text{or} \quad 0           \\
   & \mathbf{else if} (a \leq a_\mathrm{frag}):                \\
   & \qquad p = \frac{1}{2} \quad \text{or} \quad \frac{1}{4}.
 \end{split}
\end{align}
Here the first value of $p$ corresponds to the case $a\leq a_\mathrm{set}$, the second (after the ``or'') applies to sizes above $a_\mathrm{set}$. The
sizes $a_\mathrm{BT}, a_\mathrm{12}, a_\mathrm{set}$ are calculated according to
Eqs. (37), (40), and (27) in \citetalias{Birnstiel:2011ks},
\begin{align}
 a_\mathrm{set} & = \frac{2\,\alpha\,\Sigma_\mathrm{g}}{\pi\,\rhos},          \\
 a_\mathrm{BT}  & = \left[\frac{8 \Sigma_\mathrm{g}}{\pi \rhos} \cdot
 \mathrm{Re}^{-\frac{1}{4}} \cdot \sqrt{\frac{\mu \,
 m_\mathrm{p}}{3\pi \, \alpha}} \cdot \left(\frac{4\pi}{3}
 \rhos\right)^{-\frac{1}{2}}  \right]^{\frac{2}{5}},                           \\
 a_{12}         & = \frac{1}{y_a} \, \frac{2 \Sigma_\mathrm{g}}{\pi \, \rhos}
 \cdot \mathrm{Re}^{-\frac{1}{2}},
\end{align}
and the particle Reynolds number $\mathrm{Re}$ is
\begin{align}
 \mathrm{Re} & \approx \frac{\alpha \, \Sigma_\mathrm{g} \,
 \sigma_{H_2}}{2 \, \mu \, m_\mathrm{p}}.
\end{align}
Here, $m_\mathrm{p}$ is the proton mass $\mu=2.3$ the mean molecular mass in
atomic units, $y_a\simeq 1.6$ \citep{Ormel:2007bl}, $\sigma_\mathrm{H_2} \simeq
\SI{2e-15}{cm^2}$ the atomic hydrogen cross section. The fragmentation limit is
given by
\begin{equation}
 a_\mathrm{frag} = \frac{\Sigma_\mathrm{g}}{\pi\,\rhos\,b} \,\sqrt{1 - 4\,b^2},
 \label{eq:fragbarrier1}
\end{equation}
where
\begin{equation}
 b = \frac{1}{3\,\alpha}\,\left(\frac{v_\mathrm{frag}}{c_\mathrm{s}}\right)^2.
 \label{eq:fragbarrier2}
\end{equation}
If $a_\mathrm{frag}<a_{12}$, the fragmentation limit in the first turbulent regime needs to be calculated from
\begin{equation}
 \mathrm{St}_\mathrm{frag} = \mathrm{Re}^{-1/4} \, \frac{v_\mathrm{frag}}{c_\mathrm{s}}  \sqrt{\frac{2}{3 \, \alpha}}
\end{equation}
The distribution $\sigma(a)$ is normalized to the total dust surface density
\begin{equation}
 \Sigma_\mathrm{d} = \int_{-\infty}^\infty \sigma(a) \ln a.
 \label{eq:sigma-def}
\end{equation}
Under the assumption of vertically well-mixed dust (which is not applicable in
most parts of the disk), $\sigma(a)$ and $n(a)$ are directly proportional to
each other for all particle sizes. Vertical settling will reduce the vertical
scale height for larger particles. The local densities of each particle size can
be  calculated under the assumption of a settling-mixing equilibrium, as for
example in \citep{Fromang:2009ja}. An numerical implementation is included in the python module.

\section{Dependencies of DSHARP opacities on the chosen composition}
\label{sec:appendixB}

As explained in \autoref{sec:opacities}, the DSHARP opacities are based on
several approximations or assumptions, based on practical choices. As such, they
are meant to be used as a reference choice along the lines of previous
literature values, and not to be seen as the last word on the subject. To
demonstrate how some of these choices affect the resulting opacity values, we
will explore the effects of mixing rules / porosity, water abundance, and the
choice of carbonaceous material. These are however by far not the only
uncertainties. Far-infrared or (sub-)mm opacities were also found to be affected
by temperature dependencies
\citep{Boudet:2005hn,Coupeaud:2011jn,Demyk:2017bk,Demyk:2017eh}. Instead of
being compact and porous, particles could also be fractal instead
\citep{Tazaki:2016cl,Tazaki:2018cv}, and the composition and shape of the
particles are largely unknown. Exploring all these possible influences is,
however, beyond the scope of this paper, and we instead refer to dedicated
studies of this subject
\citep[for~example,][]{Draine:2006is,Kataoka:2014jk,Kataoka:2015hl,Woitke:2016gp,Min:2016hr,Tazaki:2016cl,Tazaki:2018cv}.

\begin{figure}[bt]
 \centering
 \includegraphics[width=0.5\linewidth]{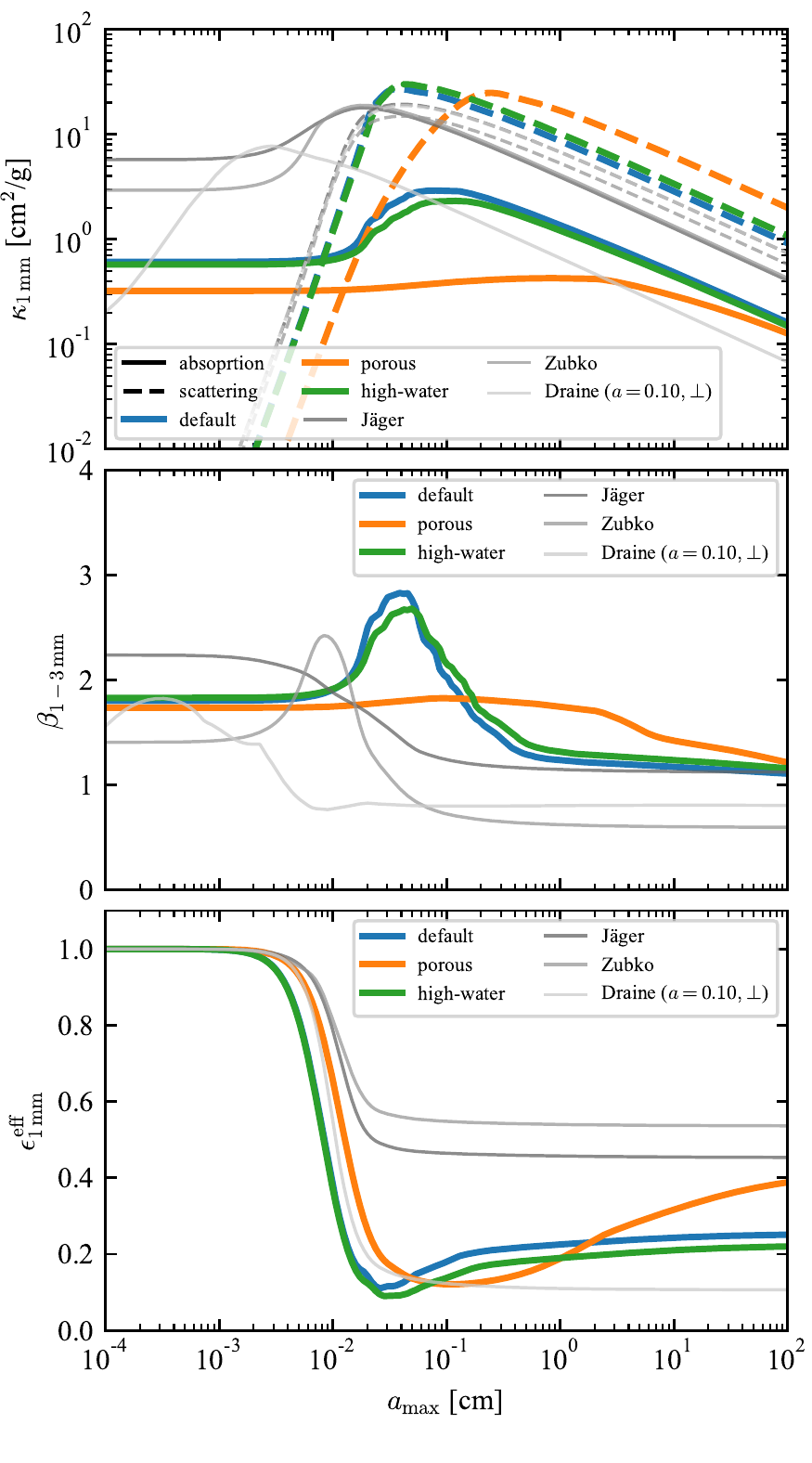}
 \caption{Optical properties for variations in the particle composition. Top
  panel: absorption (solid lines) and scattering (dashed lines) opacities for
  different grain models. Middle panel: spectral index in the 1-\SI{3}{mm}
  wavelength range. Bottom panel: absorption probability. All properties are
  calculated for a $q=3.5$ power-law size distribution with variable \amax. For
  description of the models, see \autoref{sec:appendixB}.\label{fig:appendix1}}
\end{figure}

\begin{figure*}[bt]
 \centering
 \includegraphics[width=\linewidth]{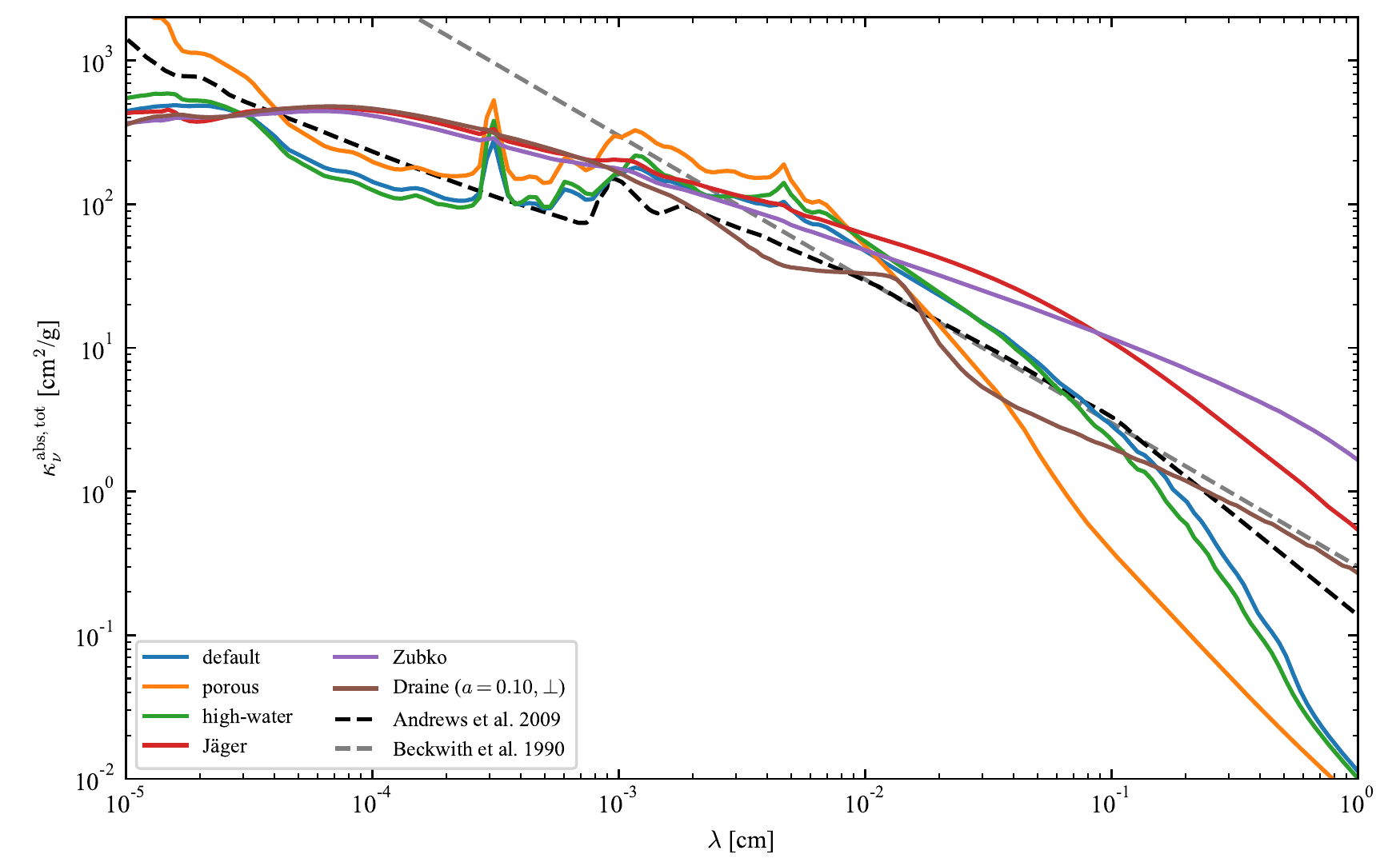}
 \caption{Wavelength dependent absorption opacity of the different grain models
  discussed in \autoref{sec:appendixB} averaged over a size distribution up to
  $\amax=\SI{1}{mm}$. The dashed lines are literature models from
  \citet{Andrews:2009jo} (black line) and \citet{Beckwith:1990hj} (grey line).
  For description of the other models, see
  \autoref{sec:appendixB}.\label{fig:appendix2}}
\end{figure*}

In the following, we will start with the DSHARP opacities (labeled as
\textit{default} in \cref{fig:appendix1,fig:appendix2}), as explained in
\autoref{sec:opacities} and then change some of those assumptions individually:
using the same relative volume fractions, we include 80\% porosity. In this
case, the Maxwell-Garnett mixing rule (\autoref{eq:maxwellgarnett}) is used and
the resulting optical properties are shown in
\cref{fig:appendix1,fig:appendix2}, labeled as \textit{porous}. It can be seen,
that in the porous case, the millimeter-range opacities are much lower, of the
order of \SI{0.3}{cm^2/g}, the highest scattering opacity is shifted to larger
\amax, and the reduced Mie-interferences also result in a flat spectral index
profile, as discussed in \citet{Kataoka:2015hl}. It should be noted that the
absorption opacity at millimeter wavelengths can be enhanced by a factor of
about 2 for silicate particles and a factor of 4 for amorphous carbon due to the
interaction of monomers, which is ignored in the Maxwell-Garnett Mie theory, as
shown in \citet{Tazaki:2018cv}.

In the next example, not the vacuum volume fraction (i.e. the porosity) is
increased, but instead the water volume fraction is raised to 60\% (labeled
\textit{high-water}). This results in very small changes in the optical
properties of the distribution including slightly increasing the water feature
around \SI{3}{\mu m}.

Very significant changes are found, if the material termed ``organics'' is
exchanged for other carbonaceous materials: as an example, we used the
carbonaceous analogue pyrolized at $T = \SI{800}{K}$ of \citet{Jager:1998vx}
(labeled \textit{Jäger}), the graphite optical constants from
\citet{Draine:2003di} (sample $a=\SI{0.01}{\mu m}$, perpendicular alignment,
labeled \textit{Draine}), and the cosmic carbon analogues (sample ACH) from
\citet{Zubko:1996wb} (labeled as \textit{Zubko}), that are also widely used in
the literature, for example in \citet{Ricci:2010gc}.
\cref{fig:appendix1,fig:appendix2} show that those carbonaceous materials cause
the strongest variations. They tend to give higher absorption opacities
\citep[see also][]{Min:2016hr}, their maximum absorption can be significantly
shifted away from $\amax \sim \lambda/(2\,\pi)$ and the resulting spectral index
is consequently affected significantly as well. Especially at millimeter
wavelength these different compounds affect the absorption opacity the most (see
\cref{fig:appendix2}).

The bottom panel of \cref{fig:appendix1} shows the absorption probability
$\epsilon_\mathrm{1 mm}^\mathrm{eff}$ (which is $1-\eta_\nu$, where $\eta_\nu$ is the single
scattering albedo) at a wavelength of $\lambda = \SI{1}{mm}$ averaged over
the MRN-like size distribution with varying $\amax$. It can be seen, that
despite the strong changes in the opacities or the spectral index, this quantity
has a very similar behavior in all cases: it is close to unity for particles smaller than
$\lambda/(2\,\pi)$ and then drops quite sharply for sizes larger than that. Only
the value reached for large \amax is very sensitive to the choices of the
opacity model. Similar to the polarization fraction of scattered thermal dust
emission discussed in \cite{Kataoka:2015hl} $\epsilon_\mathrm{eff}$ can thus be
used to constrain the maximum particle size. For the absorption and extinction
optical depths measured in \paperdsharpisella, these considerations indicate
that the particles present should be at least of a size of $\sim\SI{0.2}{mm}$.

\bibliography{bibliography}

\end{document}